# Using Technology in Digital Humanities for Learning and Knowledge Dissemination

Armanda Rodrigues, Nuno Correia
Departamento de Informática, NOVA LINCS
Faculdade de Ciências e Tecnologia, Universidade NOVA de Lisboa
{a.rodrigues, nmc}@fct.unl.pt

## Resumo

A investigação em Humanidades Digitais (DH) desenvolveu-se devido ao investimento em tecnologia para a criação de ferramentas de acesso e interação com informação de Humanidades e Património. A disponibilização dessas ferramentas diminui a distância entre os especialistas em DH e os seus estudantes a diversos níveis, não só porque estas facilitam o acesso à informação, mas também pelas tecnologias de disseminação utilizadas nessas ferramentas, destinadas à melhoria da experiência do utilizador.

A maior parte das disciplinas associadas às humanidades envolve referências geográficas e temporais, muitas vezes integradas. Essas referências têm sido cientificamente e pedagogicamente tratadas há séculos e são estabelecidas através do uso de mapas e linhas do tempo. Ambos estes suportes são hoje também disponibilizados digitalmente e o seu potencial aumentou graças à sua integração inovadora em narrativas e em mapas de narrativas, permitindo a disseminação de eventos históricos através da sua sobreposição geográfica em mapas digitais. Estas ferramentas podem ser aperfeiçoadas quando suportadas por dados ricos, como imagens, vídeos, sons e suas possíveis combinações em realidade virtual e aumentada.

Neste artigo, descrevemos um conjunto inicial de ferramentas que utilizam um subconjunto destas tecnologias e tipos de dados associados para apoiar a aprendizagem e a disseminação de dados e conhecimentos de Humanidades. Descrevemos ainda como as técnicas para disponibilização de dados e as ferramentas para melhorar a interação com esses dados podem enriquecer a experiência do utilizador e potenciar a aprendizagem e a disseminação.

**Palavras-chave:** Humanidades Digitais, Aprendizagem, Informação Geográfica, Narrativas.

## Abstract

Research on Digital Humanities (DH) has been boosted due to the investment in technology for developing access and interaction tools for handling Humanities and Heritage data. The availability of these tools lowers the distance between DH scholars and data generators, and students at various levels, not only because it facilitates access to information but also through the dissemination technologies used in these tools, designed for the improvement of user experience.

Most of the disciplines associated with the humanities involve geographical and temporal references, often integrated. These references have been scientifically and pedagogically handled for centuries and are established through the use of maps and timelines. Both these supports have been implemented and used digitally and their potential has been risen through their innovative integration with narratives, storytelling and story maps, enabling the telling of historical events in narratives superimposed on maps. These can be enhanced when supported by rich data, such as images, videos, sound, and their possible combinations in virtual and augmented reality.

In this paper, we describe an initial set of tools which use a subset of these technologies and data types to enable learning and dissemination of Humanities data and knowledge. We describe how techniques for making data available and tools for enhancing interaction with these data can improve user experience and potentiate learning and dissemination.

## Introduction

As stated by (SVENSSON, 2009), the changes experienced in the Humanities, in relation with increasing use and exploration of information technology, are no stranger to an evolution in the discipline, influencing research practices, funding, infrastructures and the emergence of networking phenomena which have further united the scientific community. Moreover, the availability of information technology tools in the field of Digital Humanities (DH) shortens the distance between its scholars and data generators, and students and other interested parties at various levels, not only because it facilitates access to information, but also through the dissemination technologies used in these tools, designed for the improvement of user experience. One evidence of this is the growth of information/knowledge repositories/websites disseminating the Humanities and their protagonists such as InPho[1], PhilPapers[2], Homo Faber Guide[3] and tDAR[4].

Most of the disciplines associated with the Humanities involve the use of geographical and temporal references, often integrated. These references have been scientifically and pedagogically handled for centuries and are established through the use of maps, timelines and other related techniques (ALLEMAN; BROPHY, 2003; PANG et al., 2017). These supports have been implemented and used digitally and their potential has been risen through their innovative integration with digital representations of narratives, using storytelling techniques and story maps, and enabling the telling of historical events in narratives superimposed on digital maps (CORREIA et al., 2005). The instantiation of these historical occurrences needs evidence support which may be realised through written accounts of the happenings, but which will gain interest from the public/learners if supported by multimedia rich data, such as images, videos or specialized audio (JAKUBOWICZ, 2007). Additionally, the increasing quality of these materials, as well as the evolution of mobile and desktop devices, currently supports their superimposition with real-time video and audio streams, enriching a visit to a relevant site with augmented reality facilities. The possibility of generating 3D models of relevant artefacts enables integration of these materials with real time feeds of the surrounding environments or immersive experiences, in virtual reality (NÓBREGA; CORREIA, 2017).

---
[1] The Internet Philosophy Ontology Project - https://www.inphoproject.org/
[2] https://philpapers.org/
[3] https://www.homofaberguide.com/
[4] The Digital Archaeological Record - https://core.tdar.org/

When developing these tools, the quality and curation of the data used is of paramount importance, ensuring the fidelity of the information conveyed to students and interested parties (POOLE, 2017). However, it is also interesting to provide some form of user data capture, to be individually added to the dataset, allowing the community to contribute to the environment, and providing a sense of belonging to a group to contributors. This capacity, possibly associated with gaming techniques, will motivate students to use the platforms and to keep coming back to them, enjoying the learning features of the study (CORREIA et al., 2005). Additionally, increased motivation to use the platforms is often related to the modelling of distinct user profiles and the availability of responsive and adaptive (or adapted) user interfaces, which will enable login and tailored use on various devices, directed at the preferred practices associated with the devices. The design of responsive interfaces should thus consider not only the device to be used, its data capacity and physical size, but also the types of possible interactions and the targeted users who will prefer the medium.

Considering the potential and current requirements of DH platforms, we present, in this paper, an initial set of platforms, using a subset of the technologies and data types already presented, created with the aim of motivating learning and dissemination of Humanities data and knowledge for diverse audiences and across various devices. Based on these experiences, we will also propose a set of functionalities and techniques which can support, in similar efforts, a successful DH technological platform, considering the skills and capacities already addressed. We will end the paper by addressing several extensions to this proposal, currently under development in the PASEV Project.

## Interactive Platforms for Knowledge Dissemination in Digital Humanities

In the following sections, we will describe a set of efforts developed for disseminating knowledge in several topics related to the Humanities, having as a primary tool, a map-based digital interactive platform. The sections are ordered chronologically, temporally organised as they were developed, and thus some evolution on the functionality and technology can be apprehended. After these descriptions, a discussion section will propose a set of capacities for Digital Platforms for the Humanities, which can potentiate and facilitate learning and dissemination of knowledge.

### LxConventos

The LxConventos platform[5] (GOUVEIA et al., 2015) was developed with the aim of reaching several types of audiences interested in the study of the evolution of the urban landscape in Lisbon, Portugal, and its transformation due to the extinction of religious orders in the 19th century. Users can view and study originally religious buildings and their changes over time, either changes to the building itself or to its surroundings. In this way, LxConventos integrates various types of multimedia data with time reference, enabling the study of the evolution of buildings from that data. Data sources include heritage data owned by the Municipality of Lisbon[6], including associated information (text, images and georeferenced drawings and videos) and geographic data layers generated for the project, which were juxtaposed on the interactive map of the city, made available on a map-based interactive interface (**Figure 1**). In this setting, the different religious buildings are represented spatially, with the possibility of viewing spatial changes to the buildings between 1834 and 2015, plus their integration with Lisbon's Historical Maps, whose transparency settings may be configured by the user. On the map, places of interest (religious buildings) are thus represented as polygons, using vector shapefiles ('Shapefiles—Portal for ArcGIS | ArcGIS Enterprise', [s.d.]).

---

[5] http://lxconventos.cm-lisboa.pt/

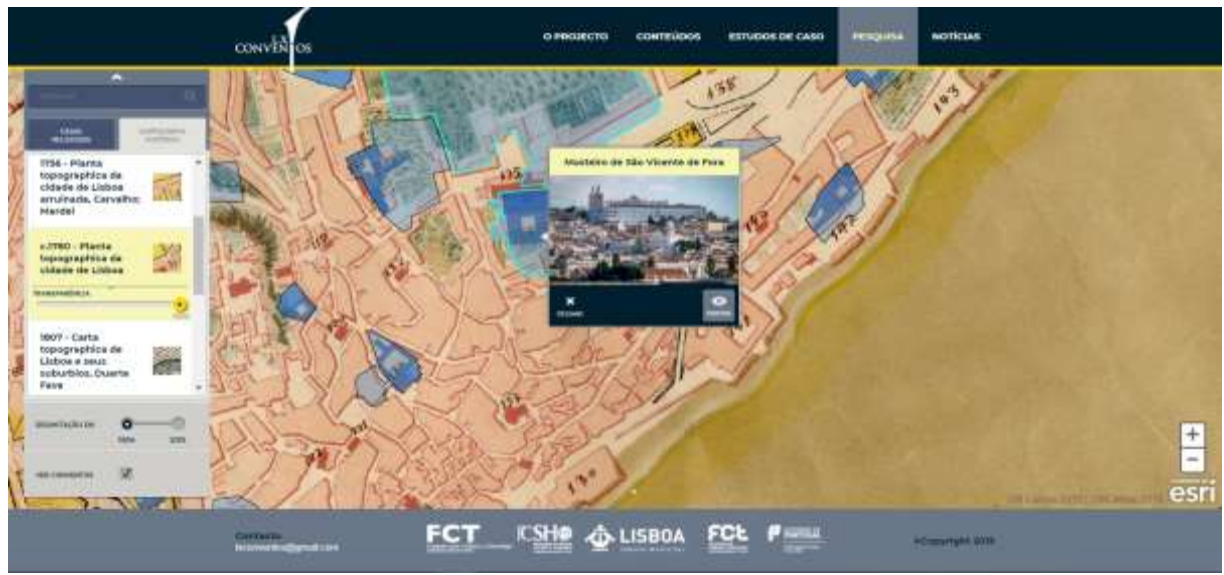

**Figure 1: LxConventos Platform – Map interactive interface**

Moreover, every relevant building was captured photographically at the time of the project (2015). The browsing of the building's evolutionary representation was made possible in perspective, due to a tool developed specifically for the project, Evolapse (BRANCO et al., 2015), to provide a chronological journey through time, supported by the available set of photographs, originating from different time periods. The developed tool provides the ability to generate visualizations of a geographic location, given a set of related images, taken at different periods in time. It automatically processes comparisons of images and establishes relationships between them. It also offers a semi-automated method to define relationships between parts of images (**Figure 2** and **Figure 3**). The LxConventos Platform is publicly available to this day, hosted by the Municipality of Lisbon.

---
[6] http://lisboaaberta.cm-lisboa.pt/index.php/pt/

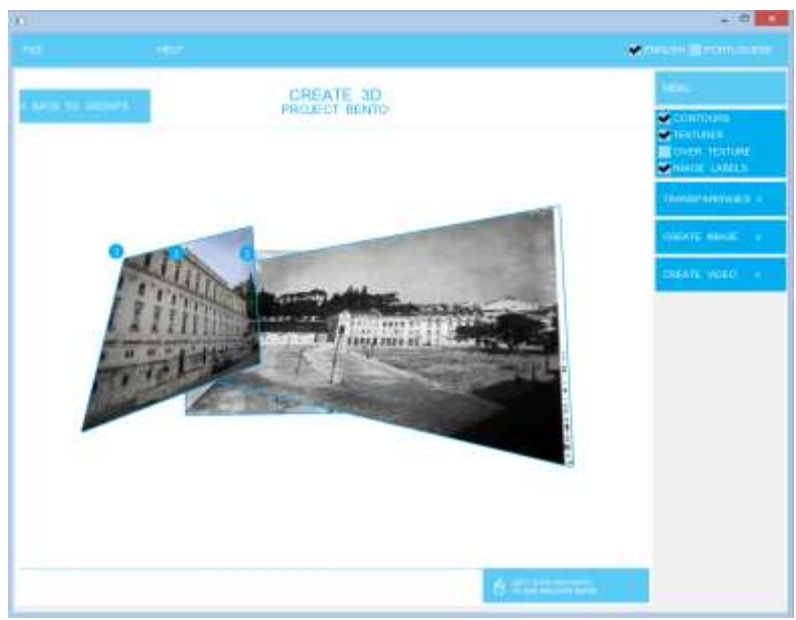

**Figure 2: Screen capture of Evolapse integration editor**

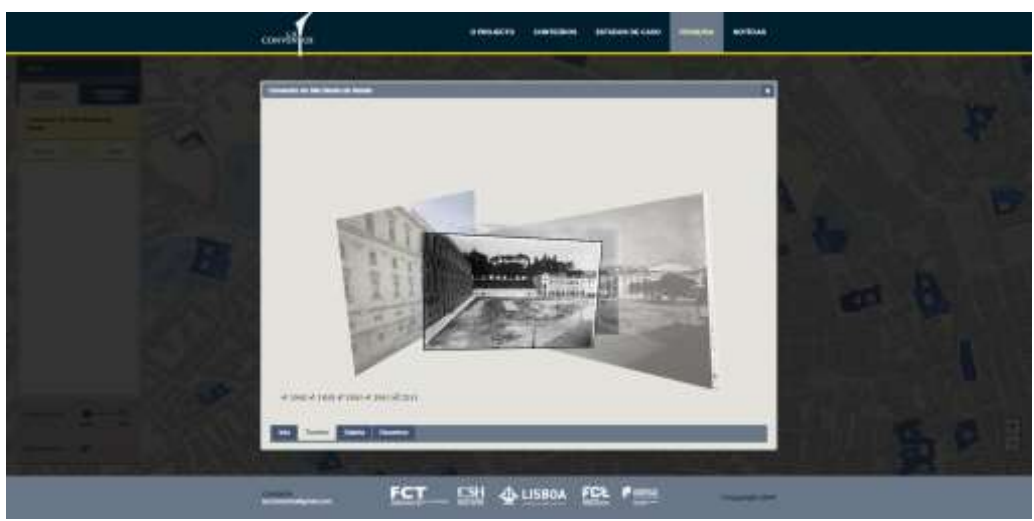

**Figure 3: Using the results of Evolapse on the LxConventos frame**

# LITESCAPE.PT

The LITESCAPE platform (ALMAS, 2015; ALMAS et al., 2015) aimed to disseminate and promote, through a responsive digital platform, the exploration of geographic locations in Continental Portugal, in relation to their literary dimension, and motivating public interaction with literary works and places mentioned in these works, through a collection of maps. An interactive and responsive web application was thus developed, allowing the visualization and manipulation of the information captured in the LITESCAPE.PT project. The application's interface used an interactive map of Lisbon, where markers referring to georeferenced literary data were overlaid, being possible to view and listen to the literary excerpts associated with the location (**Figure 4**, left). It is also possible to generate thematic geographic visualisations based on queries on literary works, authors, themes and locations (**Figure 5**). When using a mobile device, the user could click on a marker on the map and get directions to go to the place, starting on the user's current location or follow thematic routes, proposed by the application (**Figure 4**, right).

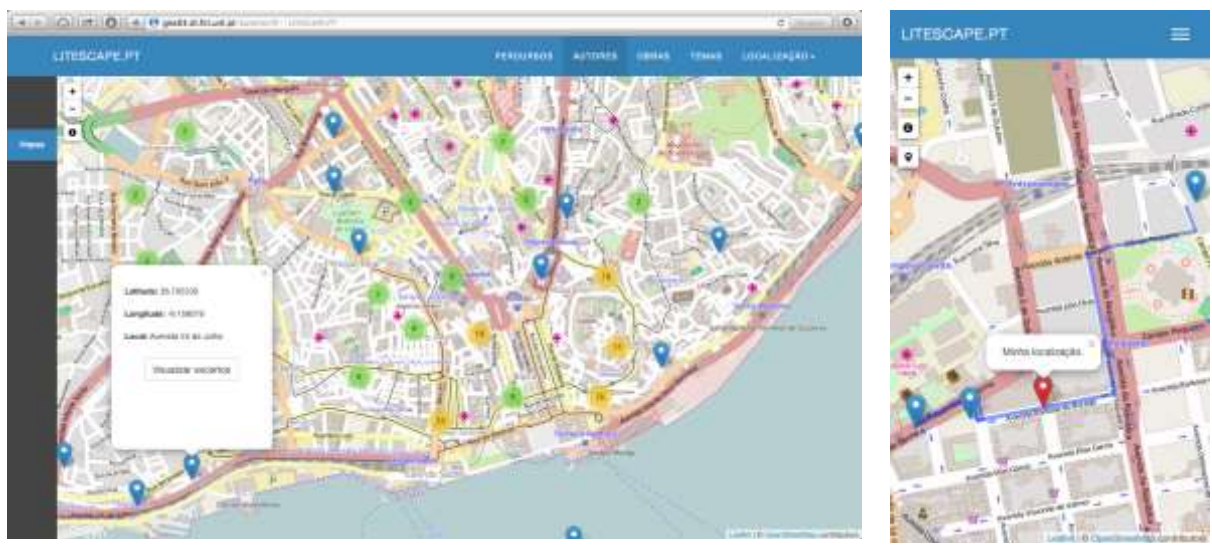

**Figure 4: The LITESCAPE.PT Platform – desktop (left) and mobile interface (right)**

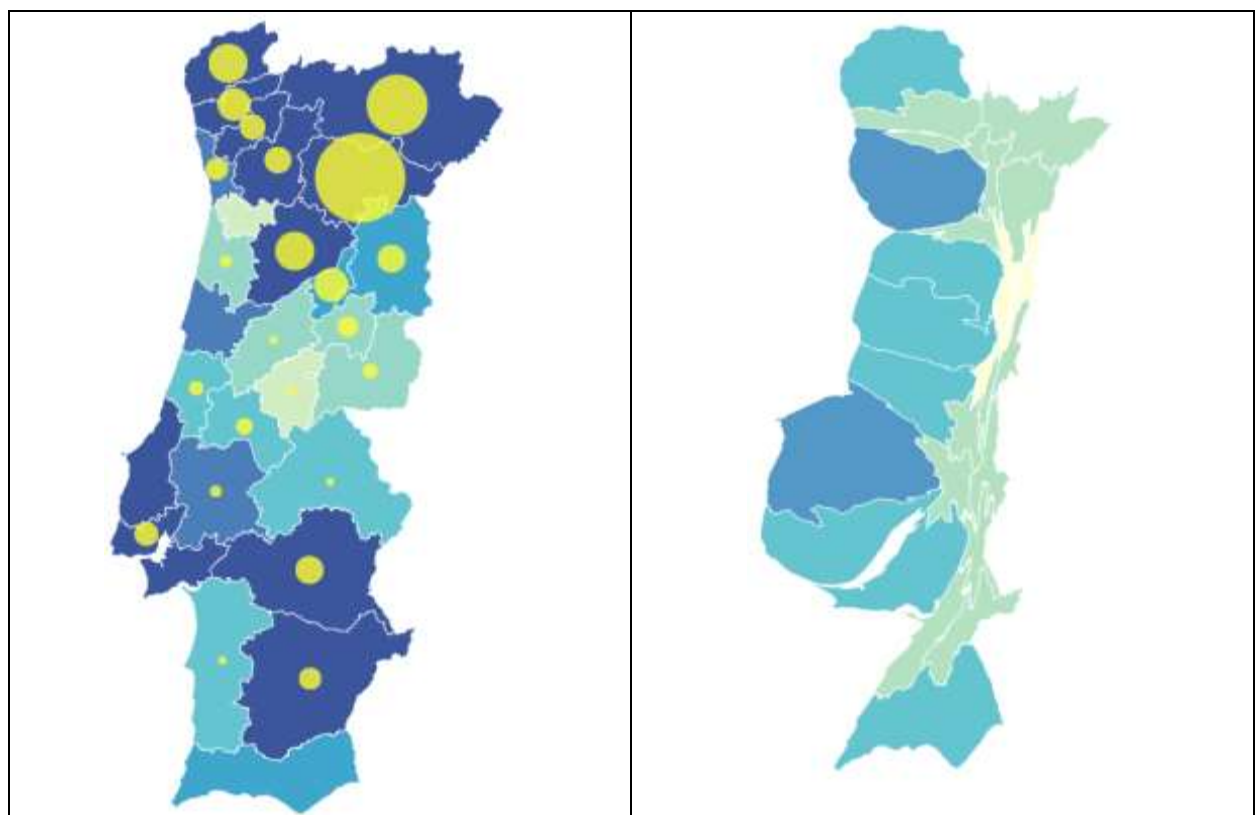

**Figure 5: Thematic visualizations generated by LITESCAPE.PT based on user searches – descriptor Wolf from theme Fauna (left) and cartogram for theme Fishing (right)**

The tool included an administration environment, enabling the coordination team to add new literary works, geo-referenced excerpts, and routes, as well as additional metadata, which made it possible to add new searches and visualizations to the platform. Multimedia data could be added to excerpts, such as images, audio, and video. Also, you could request the reading of the excerpts, improving the experience of using the app. Routes added to the platform could be organised as temporally ordered narratives of the works or as thematic geographic stories.

A two-step preliminary evaluation experiment of the application was realised with encouraging results that highlighted its contribution in improving Visual Thinking and disseminating literary, geographic and historical knowledge, while promoting visits to the places cited by the literary works.

## PASEV – Patrimonialization of Évora's Soundscape (1540 - 1910)

Project PASEV's primary objective is the appreciation and dissemination of the heritage of the auditory landscape of the city of Évora between 1540 and 1910 and the promotion of cultural tourism in the city[7]. This is a study of major interest due to the richness of the sound landscape in the municipality, covering musical activity from churches, convents, squares and theatres, among others.

In this context, the project's team is collecting diverse representations of as many historical sound events as possible in the city of Évora, related to the period between 1540 (creation of Évora's Archbishopric) and 1910 (beginning of Republic in Portugal). These testimonies are being compiled and organised in a repository, a compilation of all the audio and visual relevant documents, with the aim of enabling visitors, students and experts to experience these documents as they immerse themselves in the city. The basis for the user interaction is the interactive map of the city of Évora, associated with a timeline, where relevant and influential locations are marked, linked with visual and auditory representations (photos, drawings, sounds, music, videos), with the main focus on musical elements, which can either be represented by audio files, or written on printed scores.

Several data structures and interactive tools will be built to facilitate different experiences of these artifacts, from thematic itineraries to explore historical musical scenarios through to various virtual platforms, such as mobile apps, which may be enriched by the contribution of the visitor in loco, in the city. The platform will also integrate spatio-temporal narratives, to enable thematic travelling through the city, taking advantage of the geographic elements mentioned above. This has partially been achieved in the current prototype and will be further developed in future versions.

---

[7] https://pasev.hcommons.org/

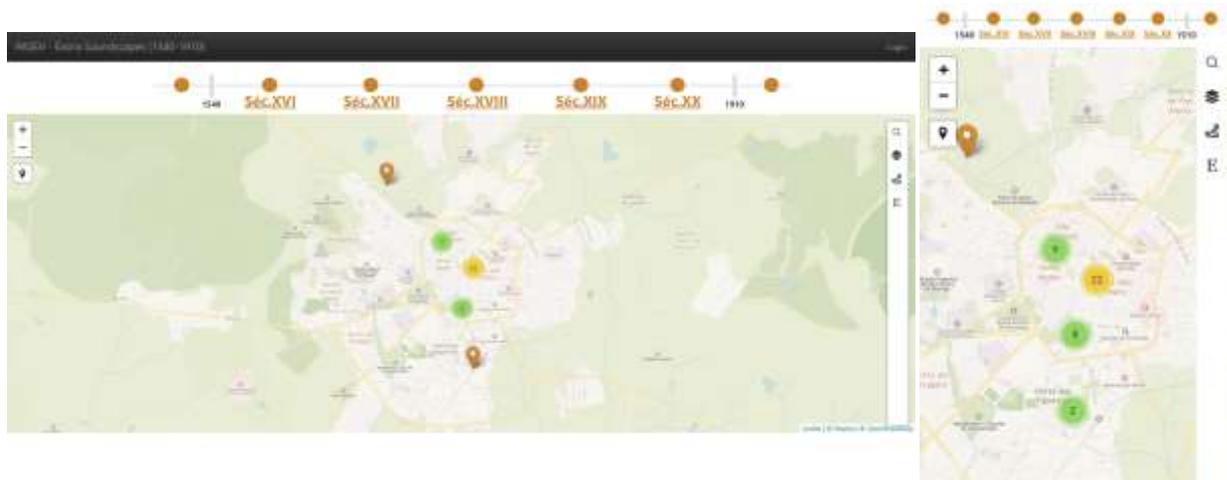

**Figure 6: The PASEV prototype platform: desktop (left) and mobile (right)**

As part of PASEV, we intend to achieve the complete realisation of the Auditory Atlas of Évora, of which the work presented in this paper is the initial step, and which will involve the capture and structuring of several types of multimedia information, as well as the development of different interfaces to present and interact with this data. At its current status, the platform presented in **Figures Figure** 6**, Figure** 7**, Figure** 8 and **Figure** 9 is an initial prototype of a responsive platform, geographically supported by the interactive map of the city of Évora, the first attempt leading to the Atlas. This prototype, which is evolving, can be used in various devices, while enabling the exploration of digital testimonies linked to the soundscape of Évora, in space and time, integrating these two dimensions, through a map and timeline interface (**Figure 6**). The platform supports several types of searches over the data, including direct queries on the map and by selection on the timeline (a seen in **Figure 6**) as well as text searches on the relevant locations and historical events organised in the repository (**Figure 7**). When interacting with a specific location, the user can access all the diverse types of data related to it, such as textual descriptions, images, video, audio and related publications (**Figure 8**).

95

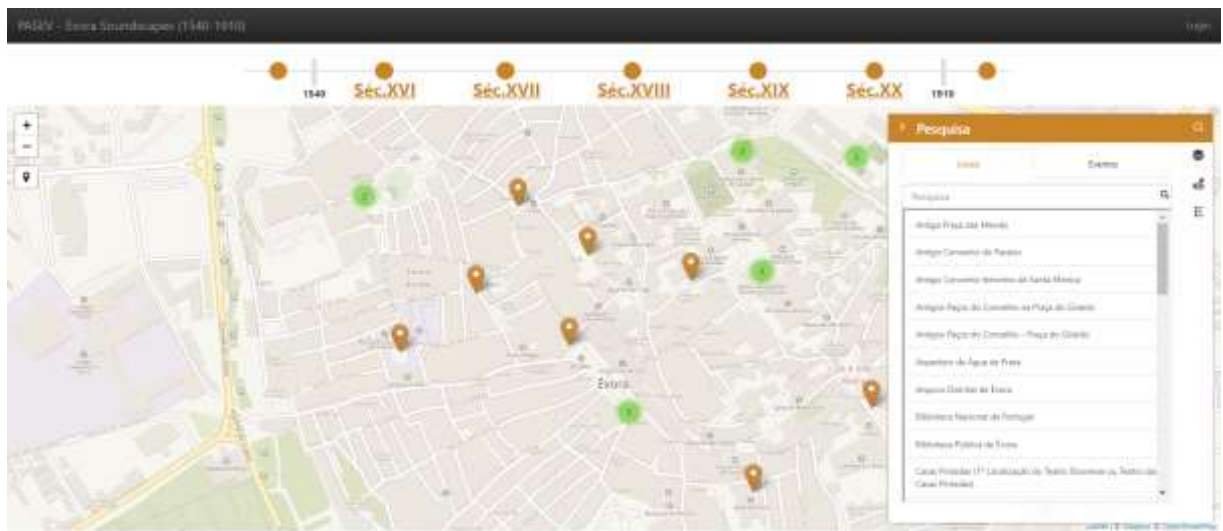

**Figure 7: PASEV search interface – the user can select locations on the map and temporally on the timeline, or use the text search interface to query locations and historical events**

Moreover, the implementation supports narratives and storytelling, allowing the representation of events that evolve spatially over time. An event is stored in the system as a temporally ordered georeferenced narrative, including all the geographic locations relevant in the context of the event, ordered temporally. However, it is more than that, as specific data can be associated with the representation of the event. In **Figure 9**, the administrator editor interface is shown, enabling the insertion of an event, which includes a geographic route. This administrator interface enables facilitated data insertion by soundscapes specialists, allowing for the growth of the database, even after the end of the project's development.

The developed platform thus allows an exploration of places regarding their musical and cultural dimension at different times, facilitating the interaction between the user and those same places (whether they are on location, using a phone, or physically absent), through the visualization and manipulation of georeferenced data sets.

The prototype has been evaluated for functionality and usability by volunteers and by expert researchers, from the PASEV project team, with positive results and is currently being upgraded in design and to correct some difficulties identified in the integration of the map and timeline interfaces (RODRIGUES; ROSÁRIO; CORREIA, 2019; ROSÁRIO, 2019). Additional interfaces are currently under development addressing 360º video and audio potential.

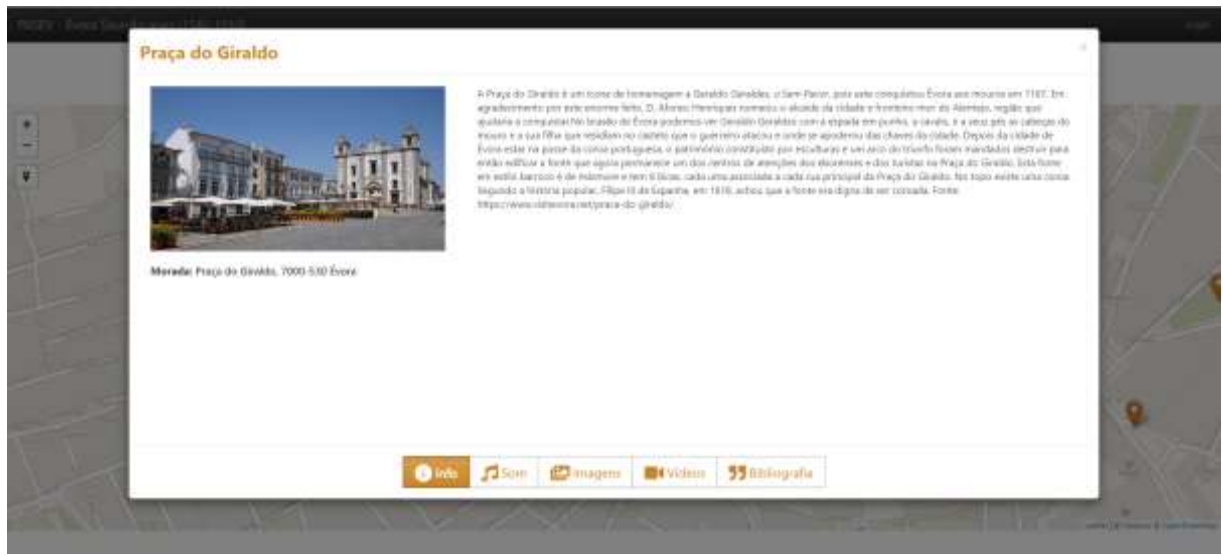

**Figure 8: PASEV data associated with each relevant location – audio, images, videos and bibliography**

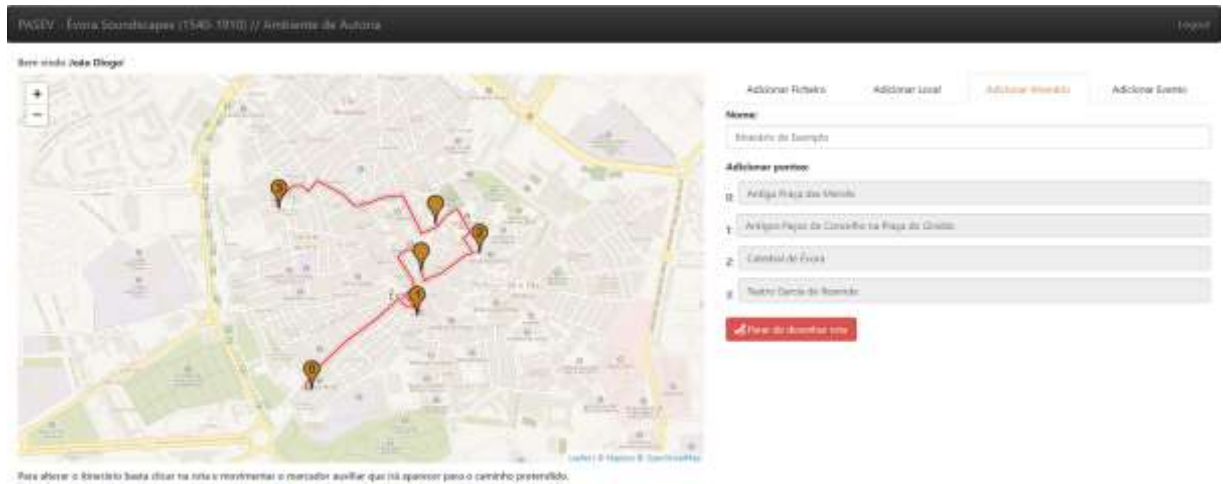

**Figure 9: The PASEV administrator interface, which enables evolution of the repository contents, including narratives associated with historic events**

## Analysis and discussion

From the platform descriptions presented before, it was possible to identify a list of functionalities and data types used to represent and disseminate knowledge in a Digital Humanities context. The completed platforms, LxConventos and LITESCAPE.PT, have both been under use for several years, which can be ascertained by the current availability of the former and by the dissemination activities proposed around the latter. PASEV is an ongoing project and, as such, its performance is being evaluated and its capacities under extension work.

**Table 1 – List of functionalities implemented in each of the presented platforms**

| Functionalities | LxConventos | LITESCAPE.PT | PASEV Prototype |
|---|---|---|---|
| Geographic Interface (Map) | √ | √ | √ |
| Timeline (or temporal dimension) | √ | √ | √ |
| Narratives | | √ | √ |
| Multimedia Data | √ | √ | √ |
|     Images | √ | √ | √ |
|     Video | √ | √ | √ |
|     Audio | | √ | √ |
| Community Contribution | √ | √ | √ |
| Data curation | √ | √ | √ |
| Responsiveness | | √ | √ |
| Mobile | | √ | √ |

From these experiences, and as presented in **Table 1**, it was possible to define a list of successful functionalities in a developmental context of a Humanities interactive digital Platform, supported by multimedia rich data and with a spatio-temporal dimension. It was thus identified the need for the integration of relevant geographic markers or clusters (associated with multimedia information at each location) juxtaposed on the base map of the region. In order to inspire trust in the information made available, curation processes are needed, which have been achieved by collaborations with domain expert teams, who are either responsible for the coordination of the projects or for content capture and management. The temporal integration or indexing of the information used in the platform has enabled linking both interfaces and the development of narratives, which can be made available to several audiences as historical events or proposed as routes for travelling and visiting. In fact, for these dimensions to be successful, the responsive and mobile attributes of the platforms must be present, to enable users to follow routes on the phones or tablets, viewing their current locations in connection with geographic tracks. The work on user directed interfaces has been less intensive but should be largely addressed in the next steps of the PASEV project.

## Concluding remarks and future work

In this paper, we have presented a body of work which has been developed in collaboration with Humanities teams, in order to create Digital Interactive platforms which can support learning and dissemination activities. The presented platforms have had successful life cycles and have contributed to the evolution of efforts that are currently under development.

As future work, in the context of the PASEV Project, we are generating specialised content for additional interfaces, such as 360º audio and video and 3D representations to be used in association with Augmented and Virtual Reality interfaces which will be integrated with the existing web responsive interfaces. These latter tools are also being extended to include improved design and narrative functionalities. Moreover, a gaming application, using the project's generated and available materials, will also be developed with the aim of addressing a younger target audience. Finally, tools for generating immersive spatial audio are being considered, in order to generate auditory content to be added to the repository.

## Acknowledgements

The authors thank João Gouveia, Fernando Branco, António Almas and João Rosário, who developed the presented platforms in the context of their MSc works. The authors also thank collaborating researchers of projects LxConventos, LITESCAPE.PT and PASEV for their contributions to the projects.

## Bibliographic References

ALLEMAN, J.; BROPHY, J. History Is Alive: Teaching Young Children about Changes over Time. **The Social Studies**, v. 94, n. 3, p. 107–110, May 2003.

ALMAS, A. M. **LITESCAPE.PT Developing a Portuguese Literary Atlas**. [s.l.] Faculdade de Ciências e Tecnologia da Universidade NOVA de Lisboa, 2015.

ALMAS, M.; RODRIGUES, A.; CORREIA, N.; QUEIROZ, A. I.; D, A. **LITESCAPE.PT - A Portuguese Literary Atlas**. Congresso de Humanidades Digitais em Portugal: Construir pontes e quebrar barreiras na era digital. **Anais**...Lisbon, Portugal: FCSH-UNL, 2015, Disponível em: <https://congressohdpt.files.wordpress.com/2015/07/livro_de_resumos.pdf>

BRANCO, F.; CORREIA, N.; RODRIGUES, A.; NÓBREGA, R. **Temporal and Spatial Evolution through Images**. 2015 IEEE International Symposium on Multimedia. **Anais**...2015

CORREIA, N.; ALVES, L.; CORREIA, H.; ROMERO, L.; MORGADO, C.; SOARES, L.; CUNHA, J. C.; ROMÃO, T.; DIAS, A. E.; JORGE, J. A. **InStory: a system for mobile information access, storytelling and gaming activities in physical spaces**. Proceedings of the 2005 ACM SIGCHI International Conference on Advances in computer entertainment technology. **Anais**...ACM, 2005Disponível em: <https://dl.acm.org/doi/pdf/10.1145/1178477.1178491>. Acesso em: 15 nov. 2020

GOUVEIA, J.; BRANCO, F.; RODRIGUES, A.; CORREIA, N. **Travelling Through Space and Time in Lisbon's Religious Buildings**. Proceedings of Digital Heritage 2015 - DH'15. **Anais**...IEEE, 2015

JAKUBOWICZ, A. Bridging the Mire between E-Research and E-Publishing for Multimedia Digital Scholarship in the Humanities and Social Sciences: An Australian Case Study. **webology**, v. 4, n. 1, 2007.

NÓBREGA, R.; CORREIA, N. Interactive 3D content insertion in images for multimedia applications. **Multimedia Tools and Applications**, 2017.

PANG, P. C. I.; BIUK-AGHAI, R. P.; YANG, M.; PANG, B. Creating realistic map-like visualisations: Results from user studies. **Journal of Visual Languages and Computing**, v. 43, p. 60–70, 1 Dec. 2017.

POOLE, A. H. "A greatly unexplored area": Digital curation and innovation in digital humanities. **Journal of the Association for Information Science and Technology**, v. 68, n. 7, p. 1772–1781, 1 Jul. 2017.

RODRIGUES, A.; ROSÁRIO, J.; CORREIA, N. **A responsive platform for the Auditory Atlas of Évora**. (V. de Sá, A. F. Conde, R. T. de Paula, Eds.)II Historical Soundscapes Meeting. **Anais**...Évora: CESEM - Universidade de Évora, 2019, Disponível em: <https://hcommons.org/deposits/objects/hc:27040/datastreams/CONTENT/content>

ROSÁRIO, J. **Uma plataforma responsiva para o Atlas Auditivo de Évora**. [s.l.] Universidade NOVA de LIsboa, 2019.

**Shapefiles—Portal for ArcGIS | ArcGIS Enterprise**. Disponível em: <https://enterprise.arcgis.com/en/portal/10.5/use/shapefiles.htm>. Acesso em: 27 jul. 2020.

SVENSSON, P. Humanities Computing as Digital Humanities. **Digital Humanities Quarterly**, v. 003, n. 3, 29 Sep. 2009.

## About the authors

| | |
|---|---|
| 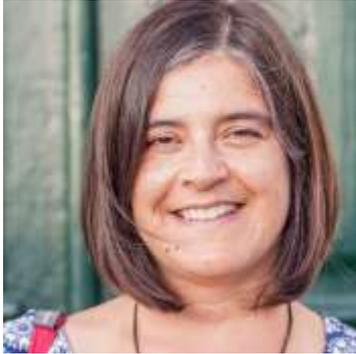 | **Armanda Rodrigues**<br><br>Armanda Rodrigues is an Associate Professor at the Computer Science Department, NOVA School of Science and Technology and an integrated member of the Multimodal Systems Group of NOVA LINCS. Armanda is interested in providing models, methods, tools and infrastructures that may enable improvements in the use of Web/Mobile GIS (Geographic Information Systems), focusing in changes in context and in collaborative environments. She has been involved in several International and national research projects related with GIS, Simulation, Web-GIS and Geo-Collaborative Systems with case studies in Emergency Management, Digital Heritage and Agronomy. Armanda Rodrigues supervised more than 30 postgraduate theses already completed and currently supervises several doctoral and master dissertations. She is the author and co-author of several GI Science and Computer Science peer reviewed publications. She also reviews and serves in the program committee of various GI national and international conferences as well as peered review journals. |
| 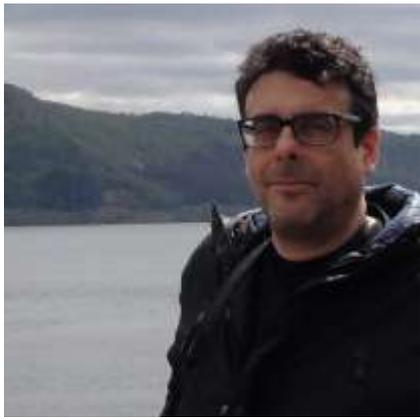 | **Nuno Correia**<br><br>Nuno Correia is a Professor at the Department of Computer Science of the Faculty of Science and Technology of the Universidade NOVA de Lisboa. He is the coordinator of the Multimodal Systems area of NOVA-LINCS, integrating a team of 11 researchers and about 20 doctoral students. His research interests cover several aspects of describing, processing, delivering and presenting multimedia information. He was a researcher at Interval Research, Palo Alto, CA, and a researcher at INESC, Portugal. He participated in several EU funded research projects and evaluated national and international projects. Current work includes video archives, mobile image processing, multitouch and pen-based interfaces for exploring art collections, cultural heritage and dance annotation. Nuno Correia supervised 12 doctoral theses and about 50 master theses already completed and currently supervises several doctoral and master dissertations. He is author or co-author of more than 100 publications in journals, conferences and books. |